\documentclass[aps,prl,preprint,superscriptaddress]{revtex4}
\usepackage[pdf]{pstricks}

\usepackage{amsmath}
\usepackage{gensymb}
\usepackage{subfig}
\usepackage{graphicx}
\usepackage{float}
\usepackage{amsthm}
\usepackage{amssymb}
\usepackage{xfrac}
\usepackage[toc,page]{appendix}
\usepackage{setspace}
\usepackage{lscape}
\usepackage[utf8]{inputenc}
\usepackage{dirtytalk}
\usepackage{tikz}

\newcommand{\hb}{\hbar}
\newcommand{\half}{\frac{1}{2}}
\newcommand{\gm}{\gamma}
\newcommand{\lb}{\lambda}

\newcommand{\eps}{\epsilon}

\newcommand{\non}{\nonumber}

\newcommand{\og}{\omega}

\newcommand{\Gm}{\Gamma}

\newcommand{\sbp}{\subparagraph*{}}
\newcommand{\comment}[1]{}
\newcommand{\Tr}{{\rm Tr}}

\newcommand{\bds}{\boldsymbol}

\newcommand{\rgl}{\rangle}
\newcommand{\lgl}{\langle}

\newcommand{\Dlt}{\Delta}

\newcommand{\alld}{\allowdisplaybreaks}

\newcommand{\expvc}[1]{\left\lgl  #1 \right\rgl}

\newcommand{\tavg}[1]{\overline{#1}}

\newcommand{\eqtion}[1]{\begin{equation} \non #1 \end{equation}}

\begin{document}
\title{Quantum Secrecy in Thermal States}
\author{Elizabeth Newton}
\author{Anne Ghesqui\`ere}
\author{Freya L. Wilson}
\author{Benjamin T. H. Varcoe}
\affiliation{Quantum Experimental Group, School of Physics and Astronomy, University of Leeds, Leeds LS2 9JT, United Kingdom}
\email[]{a.ghesquiere@leeds.ac.uk}

\author{Martin Moseley}
\affiliation{Airbus Defense \& Space}

\date{\today}

\pacs{}

\begin{abstract}
We propose to perform quantum key distribution using quantum correlations occurring within thermal states produced by low power sources such as LEDs. 
These correlations are exploited through the Hanbury Brown and Twiss effect.  
We build an optical central broadcast protocol using a superluminescent diode which allows switching between laser and thermal regimes, enabling us to provide comparable experimental key rates in both regimes. 
We provide a theoretical analysis and show that quantum secrecy is possible, even in high noise situations.
\end{abstract}

\maketitle

\section{Prelude}
\sbp  In 2016, China launched what Gibney dubbed the \textit{first quantum satellite} \cite{Gibney:2016}, the intent of which is to perform quantum key distribution between the satellite and ground stations, see for instance \cite{Liao:2017}.
This is just one of the current practical schemes designed to perform quantum secure key distribution and communication between parties; other examples include the DARPA network \cite{DARPA:2005}, the SECOQC project \cite{SECOQC:2009}, or the Durban-QuantumCity project \cite{Mirza:2010_1, Mirza:2010_2}, which use fibre-optic technologies to build quantum networks. 
These technologies rely on optical communication setups that were proven to be sufficient for performing quantum key distribution (QKD) \cite{Gisin:2002}. 
Optical setups commonly work over great distances and achieve high bit rate; for instance, the Cambridge Quantum Network achieves a secure key rate of about 2.5Mb/s \cite{Wonfor_QC:2017, Shields_QC:2017}. 
Such heavy duty infrastructure is, however, impractical for a plethora of short distance applications which nonetheless require high levels of encryption. 
Examples include key distribution and renewal between a mobile device and a medical implant, between an electronic car key and its lock or even between a mobile device and a password blackbox. 
These low power applications may need shorter key, lower bandwidth and as a result, an infrastructure built on high power lasers, single photons or entangled photons sources, may well be unsuitable. 
Reducing light source requirements to LEDs producing thermal states would allow us to explore the realm of low power applications and to appeal to a different set of customers.

The consideration of thermal states as a resource for QKD is not merely a technological preference, quite far from it. 
Thermal radiation is bunched, meaning that its quanta are likely to be detected in correlated pairs.
These correlations produce quantum discord, as demonstrated by Ragy and Adesso using the R\'enyi entropy \cite{Ragy:2013}.
Furthermore, Pirandola \cite{Pirandola:2014} establishes theoretically that non-zero quantum discord is necessary for QKD, and that positive discord in a central broadcast-type protocol allows a quantum secure key to be extracted even with high levels of noise. 
Indeed, quantum discord has been established as a measure of quantum correlations \cite{Modi:2012}.
Correlations can be qualified using the second-order correlation coefficient, generally known as $g^{(2)}(\tau)$, defined as \eqtion{g^{(2)} (\tau) = \frac{\expvc{Y(t) Y(t+\tau)}}{\expvc{Y(t)} \expvc{Y(t+\tau)}}\,,} where $Y(t)$ is the radiation intensity. 
Radiation can then be characterised using $g^{(2)}$ as: anti-bunched (purely non classical) when $g^{(2)}(0) < 1$,  coherent when $g^{(2)}(0) = 1$, and  bunched when $g^{(2)}(0) > 1$ \cite{Fox:2006}. 
We use this classification in order to experimentally verify that we are operating in the thermal regime, as can be seen later in Figure~\ref{g2coef}. 

To exploit the correlations within bunched pairs requires the photon pairs to be separated and shared between two parties (e.g. Alice and Bob); this is done using the Hanbury Brown and Twiss (HBT) interferometer \cite{HBT:1956_1, HBT:1956_2}, designed in the 50's to remedy the shortcomings of amplitude interferometers, such as the Michelson interferometer \cite{Michelson:1881, Michelson:1887}, used in astronomy to determine the radius of stellar objects. 
The HBT's table-top set-up is simple: a source shines onto a beamsplitter, creating two arms, each shining onto a separate detector. 
The theory behind the observed interference effect has been studied by a number of authors, amongst the first Purcell \cite{Purcell:1956} and Mandel \cite{Mandel:1958, Mandel:1959}, whose papers provide a nicely intuitive pre quantum optical description (see \cite{GlauberQO:63}) of the statistics. 

Mandel's analysis relies on the fact that what we count are not the photons themselves, but the photoelectrons ejected by the detector. 
Fluctuations in that number have two origins: very fast fluctuations in the intensity of the incoming signal, and the stochasticity of the reaction of the photo-sensitive material to its interaction with the radiation field (namely the ejection of a photoelectron).
However, the detector is also fundamentally limited by its reaction time (or bandwidth) and therefore, very fast fluctuations may occur undetected. 
Yet, if these fluctuations are invisible, the correlations between them are not; in the original experiment and in current radio astronomy, the data used for calculations is often data which has been fed through a correlator (either hardware or software). 

The production of photoelectrons is completely characterised by its average number $\tavg{n_T}$ and several experiments have been performed to estimate it, such as \cite{Tan:2014, Martinez:2007, Koczyk:1995}. 
These photon-counting experiments highlight that the bandwidth of the detector relates to the coherence time $\tau_c$ of the source, and that when the observation time $T\ll \tau_c$, $\tavg{n_T}$ follows a Bose-Einstein distribution. 
This is in fact how thermal states are usually modelled in quantum optics, especially since this distribution naturally arises through the modelling of blackbody radiation. 
However, the distribution is also generally used without acknowledgments of its caveats. 
If the source's coherence time is suitably long (i.e. the source has a linewidth in the order of kHz), then a detector bandwidth in GHz will allow for the use of the Bose-Einstein distribution, and the resolving of thermal behaviour. 
However, if the laser linewidth approaches 1Hz, a GHz bandwidth is simply too wide to observe thermal behaviour; all the relevant correlations will be lost. 
Such a caveat was neglected in articles such as \cite{Weedbrookpra:2012} and contributes to limited conclusions.

A natural objection to the use of thermal states for quantum cryptography is the lifetime of the correlations, perhaps in light of the fragility of entanglement. 
However, current common implementations include very large telescope arrays such as VLA, ATCA or soon the SKA; another famous usage is the observation of the cosmic microwave background. 
Furthermore, the original HBT experiment was performed in the optical regime.
In either regime, the correlations literally survive astronomical distances in free space. 
The use of thermal states therefore, naturally emerges as a potential partner to optical QKD techniques, especially with the rise of technologies such as WiFi and Bluetooth, which offer ever-increasing possibilities, such as through wall or medium range free space communications.
Furthermore, we will show in the following, that in either the optical or the microwave regime, the protocol described below is quantum secure. 

\sbp Next we describe the protocol we propose. 
We continue with a discussion of the eavesdropper (Eve), which will naturally lead us to analysing the security of the protocol and its theoretical modelling.
As such, we show experimental results, as well as further theoretical discussions, including on the issue of detector noise.

\section{Protocol}

\begin{figure}
    \centering 
     \includegraphics[scale=0.5]{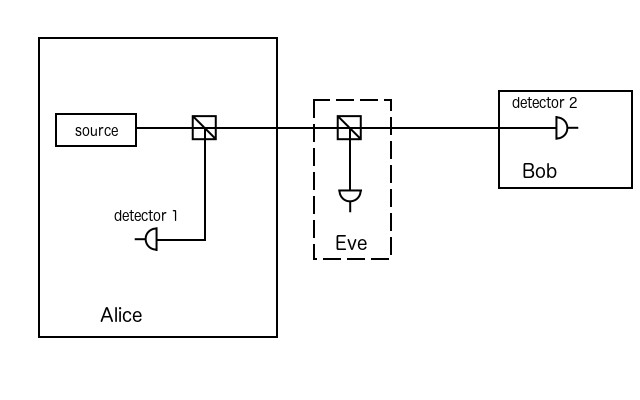} 
\caption{(Colour online) Schematic of the protocol. Even though Alice controls the source, this is a central broadcast because the signal is split between Alice and Bob. Alice does not prepare their state.}
   \label{theosetup}
\end{figure}

\begin{figure}
    \centering 
     \includegraphics[scale=0.95]{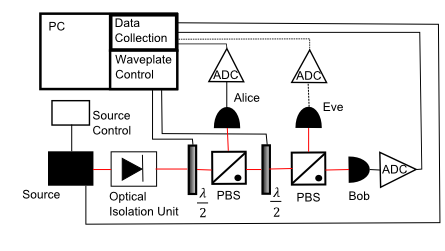} 
\caption{(Colour online) Diagram of the experimental set-up. Thermal or coherent light is produced at the source. The combination of half wave plate ($\dfrac{\lambda}{2}$) and polarising beam splitter (PBS) acts as a controllable beam splitter, allowing a controlled amount of light through. This first combination acts as $\eta_1$, directing part of the beam to Alice, and the second combination acts as $\eta_2$, directing a further part of the beam to Eve.}
   \label{protocol_setup1}
\end{figure}

We propose a central broadcast protocol, reminiscent of Maurer and Wolf's scenario 1 in \cite{Maurer:1999}, shown on Figure~\ref{theosetup}.
It is described as follows :
\begin{itemize}
\item Alice creates a beam from a trusted thermal source. 
\item They then use a trusted beamsplitter with transmittance $\eta_1$ to divert and detect part of the transmission and send the rest on to Bob (Eve). 
\item The bunched nature of a thermal source means that fluctuations present at Alice's detector are correlated with those at Bob's detector. 
\item These fluctuations can be sliced into bits any number of ways, but with no loss of generality, we assume here that a fluctuation above the signal mean is a 1 and one below the mean is a 0. 
\item In order to detect an eavesdropper, Alice sends small random chunks of data to Bob who performs a $g^{(2)}$ calculation to verify thermality.
\item Alice and Bob now have a stream of independent and randomly correlated bits from which they can derive a key, the security of which they can improve with Cascade and Advantage Distillation, as per any QKD scheme. 
\end{itemize}
This scheme was implemented as shown on Figure~\ref{protocol_setup1}.
In order to simulate high levels of noise, we consider an attenuator channel between $\eta_2$ and Bob, equivalent to adding a beamsplitter of transmittance $\eta_4$ between $\eta_2$ and Bob, with a input state of variance N at the second input arm.

Let us emphasise that this is not a prepare-and-send scheme, but instead relies on central broadcasting. 
This means that although we assume that Alice controls the source, they do not in fact, prepare their states. 

Even if Eve interferes with the signal on its way to Bob via the most powerful attack available, we assume that she has no control over any part of Alice's apparatus, including the source, the beamsplitter ($\eta_1$) or the detector.  
Similarly, she has no control over Bob's detector. 

The security of this protocol arises from the quantum correlations within the thermal fluctuations, those responsible for the Hanbury Brown and Twiss effect. 
Upon arrival at $\eta_1$ from the source, a bunched pair will either travel whole to Alice, travel whole onwards to face $\eta_2$ or split between Alice and $\eta_2$. 
Any pair travelling onwards to $\eta_2$ (and then $\eta_4$) will suffer the same fate, but the pairs we can exploit are those splitting between Alice and either Bob or Eve. 
Using the quantum correlations within the pairs is what allows us to cascade the beamsplitters this way, however, at the cost of photons pairs, and so of correlations. 
For instance, when $\eta_1$ is at 50\%, half the light goes to Alice, the rest is to be shared between Eve and Bob.
If $\eta_2$ is also at 50\%, only a quarter of the original source signal is available to Bob (provided full transmission at $\eta_4$).

\sbp To model our protocol, we must know how to model Eve and for that, we must understand what actions she can take. 
Alice controls everything, from the source to their detector, including $\eta_1$.
This means that the only place for Eve to ``insert'' herself is on Bob's branch of the distribution, much like on most current prepare-and-send QKD schemes.

By current standards, the most powerful attack at Eve's disposal is a collective Gaussian attack.
This is typically modelled by Eve mixing one mode out of an EPR sate with Bob's signal and recording the outcome for measurement upon Alice and Bob's classical communication \cite{Grosshans:2007}. 
As such, it is the attack modelled in this paper.

However, it is doubtful that Eve can, in fact, obtain any relevant information in a central broadcast scheme (CBS).
The reason for that lies in the physics of the correlations themselves.

One conceivable attack would be for Eve to measure the signal going to Bob and reproduce it.
We rely on bunched pairs, correlated via second-order temporal coherence.
Should Eve be able to measure and reproduce Bob's photons fast enough, what prevents her from escaping detection?

The operative words in that sentence are ``fast enough''.
Eve cannot beat Heisenberg's uncertainty principle, which limits her ability to detect and recreate withing a specific time, which is the signal coherence time $\tau_c$. 
In the case of our experiment, $\tau_c = 1 \mu s$. 
Correlations in bunched pairs exist only for detection during that time. 
Simply applying Heisenberg's uncertainty principle, setting $\Dlt t = \tau_c/2$, yields $\Dlt E \geq 6.6 \cdot 10^{-10} eV$ for a photon of energy $E = 1.59 eV$.
This is the maximum uncertainty which Eve is allowed for detecting and recreating the photon, to have a chance at fooling Alice and Bob by making Bob's detector click within the allotted time. 
This uncertainty does not of course, account for shot noise, and Eve can only allow for detection and/or preparation noise within the $\Dlt E$ margin.
Vacuum energy here is $E_0 = \half \hb c / \lb \approx 0.13eV$; this means that a single inescapable unit of shot noise is enough to push Eve past her limit. 

This is a very crude argument, but it demonstrates that a simple intercept-and-resend scenario is useless.
Actually, as we have explained before, Eve gains very little in using even an entangling cloner, because of the probabilistic ways that the photon pairs will split at $\eta_1$ then at $\eta_2$.

\section{Modelling}
\sbp Thermal states are Gaussian states; these states can be easily defined and manipulated through their first and second moments \cite{Eisert:2003, GarciaPatron:2007}. 
The former are contained in the displacement vector $\expvc{\hat{r}}$, where $\hat{r}$ is the system's operator, and $\rho$ the state's density operator. 
The second moments are contained in the covariance matrix $\gm$ defined as \eqtion{\gm_{ij} = \Tr \left[ \rho \left\{ (\hat{r}_i - \expvc{\hat{r}_i}),(\hat{r}_j - \expvc{\hat{r}_j})\right\}\rho \right]\,,} where we write the anti-commutator using $\left\{ \right\}$. 

A thermal state has covariance matrix $\gm_{in} = 2(\bar{n} + 1) \bds{I} $, where $\bar{n}$ is the average photon number and $\bds{I}$ the identity matrix, and null displacement. 
We consider in the present work, a displaced thermal state, with covariance matrix as before, but with non-null displacement (it can also be construed as a noisy coherent state). 

We use the Bose Einstein distribution \begin{equation} \label{BEdist} \bar{n} = \frac{1}{e^{\sfrac{\hb \og}{k_B T}}-1}\,,\end{equation} acknowledging all the caveats highlighted in the introduction, and consider narrowband detectors measuring radiation at $30GHz$ and $T=300K$, so that $\bar{n} = 1309$. 

A beamsplitter is modelled as \eqtion{\bds{V}_i = \left( \begin{array}{cc} \sqrt{\eta_i} \bds{I} & \mu_i \bds{I} \\ -\mu_i \bds{I} & \sqrt{\eta_i} \bds{I} \end{array} \right) \,,}
where $\mu_i = \sqrt{1 - \eta_i}$ represents the loss.

The input state at the first beamsplitter contains the thermal source and a vacuum state; it has covariance matrix and operator vector
\eqtion{\gm_{in}  = \left(\begin{array}{cc} V_s^x & 0\\ 0 & V_s^p \end{array} \right)\bigoplus \bds{I}  \, .} 

Since we give Eve an entangling cloner, she inputs one mode of her state at $\eta_2$ so the input state is of the form \eqtion{\gm_{in}^{\eta_2}  = \gm_{out}^{\eta_1} \bigoplus \left(\begin{array}{cc} V_e^x & 0 \\ 0 & V_e^p \end{array} \right)\,.}

In fact, Eve's full input state can be written as \eqtion{\gm_{eve} = \left( \begin{array}{cc} \nu \bds{I} & \sqrt{\nu^2 - 1}\bds{Z} \\ \sqrt{\nu^2 - 1}\bds{Z} & \nu \bds{I}\end{array} \right)\,,} with $\bds{Z}$ the Pauli-Z matrix.
However, only one mode of the EPR mixes with the legal signal at $\eta_2$. 
Since the rest of their state is unavailable to us, and of little practical value, we can trace it out for the sake of clarity. 
We can make further assumptions on Eve's state; she can be merely there and tap the channel, in which case, her inputs is one shot noise unit $V_e = 1SNU$.
If she inputs a state, the minimum variance she can get away with is $V_e = 2SNU$, where 1SNU comes from her coherent state and the complementary 1SNU through shot noise.

We make the channel between $\eta_2$ and Bob a thermal noise channel by inputting a state of variance
\eqtion{N = \frac{\eta_4 \chi}{1-\eta_4} \,, \qquad \text{with} \qquad\chi = \frac{1-\eta_4}{\eta_4} + \eps \,,} and $\eps$ the channel excess noise \cite{GarciaPatron:2007}. 
The input state at $\eta_4$ is \eqtion{\gm_{int}^{\eta_4}= \gm_{out}^{\eta_2} \bigoplus \left(\begin{array}{cc} N & 0\\ 0 & N \end{array} \right) \,,}
where $\gm_{out}^{\eta_2} $ is the state at the output of $\eta_2$ and $N$ as defined previously.

The output covariance matrix is
\eqtion{\Gm_{out}= \left(\begin{array}{ccccc} \Gm_{a} & \Gm_{ea} & \Gm_{ab}   & \Gm_{an} \\ \Gm_{ea}  & \Gm_e  & \Gm_{eb}  & \Gm_{en}    \\ \Gm_{ab} & \Gm_{eb}   & \Gm_b  &\Gm_{bn} \\ \Gm_{an} & \Gm_{en}  & \Gm_{bn}  & \Gm_{n}  \end{array}\right)} 
where the sub-matrices of interest are
{\alld
\begin{align}\Gm_a =& \left( \begin{array}{cc} \mu_1^2 V_s^x + \eta_1 & 0 \\ 0 &\mu_1^2 V_s^p + \eta_1 \end{array} \right)\,, \quad \Gm_e = \left( \begin{array}{cc} \mu_2^2 (\eta_1 V_s^x + \mu_1^2) + \eta_2 V_e^x & 0 \\ 0 & \mu_2^2 (\eta_1 V_s^p + \mu_1^2) + \eta_2 V_e^p  \end{array} \right)\,, \non
\\ \Gm_{ea} =& \left( \begin{array}{cc} \mu_1 \sqrt{\eta_1} \mu_2 (V_s^x-1) & 0 \\ 0 & \mu_1 \sqrt{\eta_1} \mu_2  (V_s^p-1)\end{array} \right)\,, \non
\\  \Gm_{eb} =& \left(\begin{array}{cc} - \mu_2 \sqrt{\eta_2} \sqrt{\eta_4} (\eta_1 V_s^x + \mu_1^2 -V_e^x) & 0 \\ 0 & - \mu_2 \sqrt{\eta_2} \sqrt{\eta_4} (\eta_1 V_s^p + \mu_1^2-V_e^p  ) \end{array} \right)\,, \non
\\ \Gm_b =& \left( \begin{array}{cc} \eta_4 \left(\eta_2 (\eta_1 V_s^x + \mu_1^2) + \mu_2^2V_e^x  \right) + \mu_4^2 N& 0 \\ 0 &  \eta_4 \left(\eta_2 (\eta_1 V_s^p + \mu_1^2) + \mu_2^2 V_e^p  \right)+ \mu_4^2 N \end{array} \right) \,, 
\end{align}}

The remaining block sub-matrices are given in the Appendix.

The secrecy will be witnessed using the secret key rate $K(A:B\parallel E)$, defined here in terms of its lower bound as $K(A:B\parallel E) =  I(A:B) - \chi(B:E)$, where $\chi(B:E)$ is the Holevo bound, between Bob and Eve, which maximises their mutual information $I(B:E)$.

We define the mutual information as 
\eqtion{ I(A:B) = \, S(\Gm_a) + S(\Gm_b) - S(\Gm_{ab})\,,}
where $S(\Gm)$ is the Von Neumann entropy.
The Von Neumann entropy $S(\rho) = -\Tr \left(\rho \log \rho \right)$, for a Gaussian state, is simply determined in terms of the symplectic eigenvalues $x_i$ of its covariance matrix $\Gm$ \cite{GarciaPatron:2007}, as $S(\Gm) = \sum_{i=1}^k g(x_i)$, where \eqtion{g(x) =  \left(\frac{x+1}{2}\right)\log\left(\frac{x+1}{2}\right) - \left(\frac{x-1}{2}\right)\log\left(\frac{x-1}{2}\right)\,.} 

The correlation in the pairs shared between Alice and Bob is described with the quantum discord, $D(B|A)$, defined as the difference between the mutual information $I(A:B)$ and the classical mutual information $J(B|A)$ (or $J(A|B)$). 
$I(A:B)$ quantifies all possible correlations between Alice and Bob, but $J(B|A)$ quantifies those measured by local operations at Alice's and Bob's sites. 
We define the discord $D(B|A)$ as,
\begin{align}  D(B|A) =& \, S(\Gm_a) - S(\Gm_{ab}) +\min_{\Gm_0} S(\Gm_{b|x_A}) \non \\ \text{where} \qquad \Gm_{b|x_A} =& \, \Gm_b - \Gm_{ab}(X\Gm_a X)^{-1} \Gm_{ab}^T \non  
\end{align}
is the covariance matrix of B conditionned by a homodyne measurement on A \cite{Weedbrookrmp:2012}, with $X = \left(\begin{array}{cc} 1& 0 \\ 0 & 0 \end{array}\right)$ and $()^{-1}$ the pseudo-inverse.
We write the Holevo bound $\chi(B:E)$ as
\begin{align}\chi(B:E) = S(\Gm_e) - S(\Gm_{e|x_b})  \,, \non \end{align}
where we estimate $\Gm_{e|x_b}$ as above, using the submatrices we have derived through our modelling.

\section{Results}
\sbp The protocol was realised experimentally, using a tuneable laser consisting of a superluminescent diode and an external cavity. When run above the operating current, the laser emitted non-thermal coherent light, and when run below this operating current, it acted as a diode, producing thermal light. 
This effectively acts as a switch allowing the addition or removal of thermality in the source without altering any other part. We apply no modulation to the signal beyond that of the source. 
Just as shown on Figure~\ref{protocol_setup1}, the source diode shines onto a variable beamsplitter.
The various signals are detected by photodiodes and the signals then fed to an oscilloscope. 
The superluminescent diode has a typical lasing linewidth of 100kHz, the photodiodes' bandwidth is 100MHz and the oscilloscope's 150MHz.

\begin{figure}
    \centering
     \includegraphics[scale=0.3]{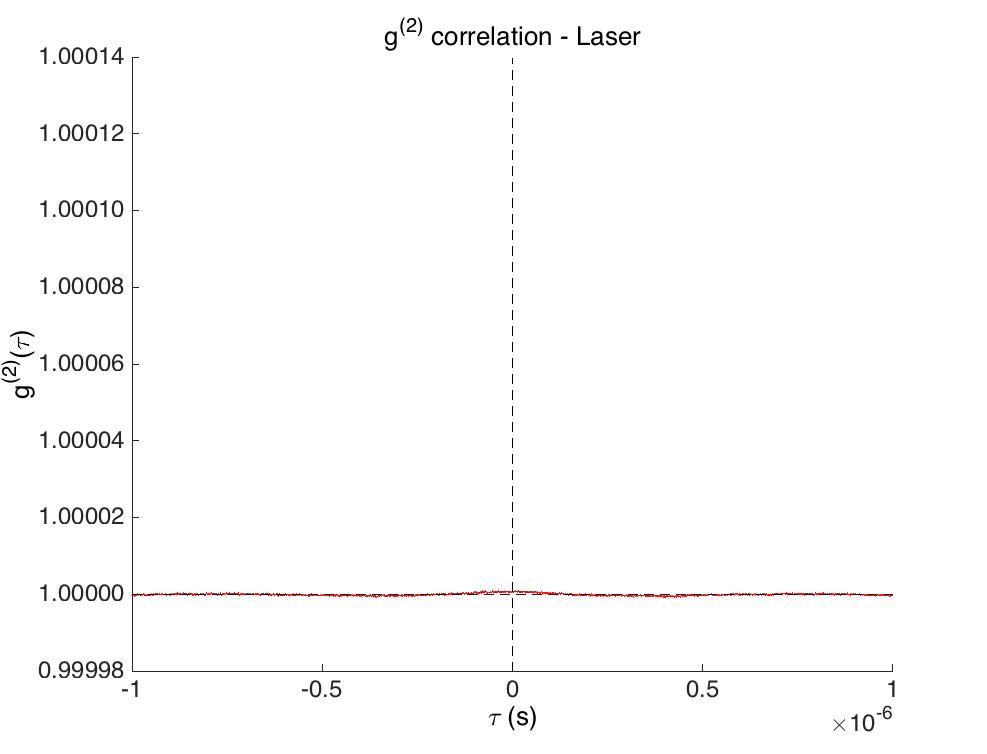} 
    \caption{(Colour online) Experimental results : second order correlation coefficient for coherent states.}
     \label{g2coefcoh}
\end{figure}

\begin{figure}
	\centering
	\includegraphics[scale=0.3]{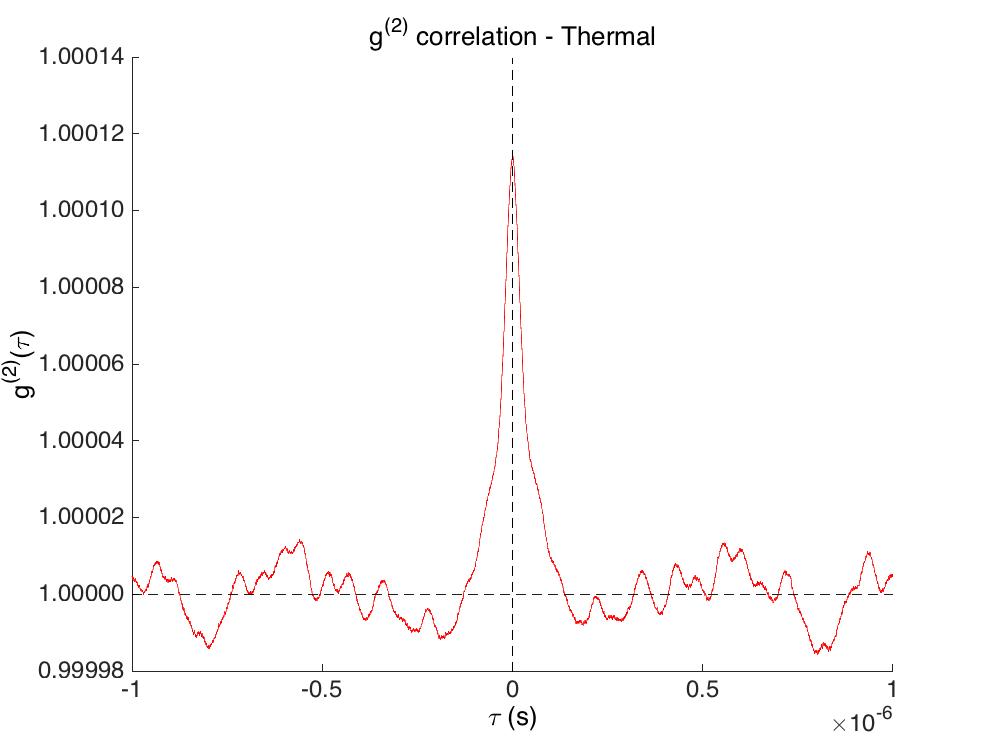}
	\caption{(Colour online) Experimental results : second order correlation coefficient for thermal states.}
	\label{g2coefth}
\end{figure}

This thermality is seen in Figures~\ref{g2coefcoh} and \ref{g2coefth} where the second order correlations at each regime are shown. 
We can easily see that \eqtion{g^{(2)}(0)|_{thermal} > g^{(2)}(0)|_{coherent}\,,} which we expected.
Figure~\ref{g2coefth} shows that in the thermal regime, $g^{(2)}(0)>1$, which means that we are dealing with bunched radiation, fluctuations and as a result, correlations. 
Due to the relatively slow photodiode bandwidth in comparison to the linewidth of the laser, the second order correlations observed will be considerably less than 2. We calculate them to be of the order 1.0001 \cite{Loudon:2000}, which is in agreement with the experimental results.
Furthermore, we see that since $g^{(2)}|_{coherent} \approx 1$, there are no correlations in coherent light because there are no fluctuations, as we expect theoretically.
Now that we have established that we have a thermal source, we shine it through two variable beamsplitters, the first dividing a portion to Alice, and the second splitting the remaining light between Eve and Bob.

\begin{figure}
    \centering
     \includegraphics[scale=0.5]{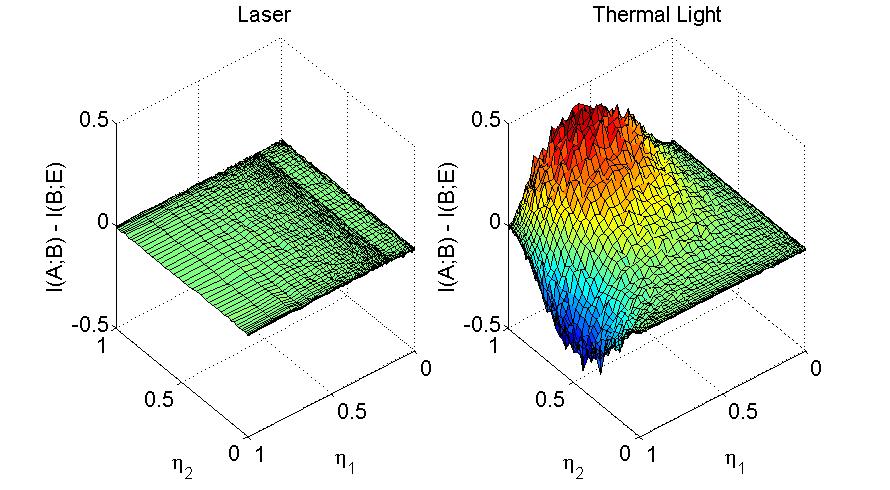} 

    \caption{(Colour online) Experimental results : secret key rate for coherent states (left) versus thermal states (right). $\eta_1$ is controlled by Alice; when $\eta_1 = 1$, she has no signal and therefore, the key rate becomes negative. Similarly, as $\eta_1 \rightarrow 0$, she gains the advantage over Bob and Eve. As expected, the most advantageous value is when $\eta_1 = 0.5$. }
   \label{thervscoh}
\end{figure}

We plot the secret key rate $K'(A:B\parallel E) = I(A:B) - I(B:E)$ in Figure~\ref{thervscoh}.



The use of thermal light clearly provides a higher key rate than coherent light. 
The most optimal regime is for Alice and Bob to both receive equal portions of light, so the key rate peaks when $\eta_1 = 0.5$. 
We note easily that $K'(A:B\parallel E)$ (which is a lower bound) is negative when $\eta_1 \rightarrow \, 1$, so when no signal goes to Alice and as a result, they share no information with Bob. 
Recall that Alice is in control of $\eta_1$, and we assume that Eve cannot adversely affect this beamsplitter. 
When $\eta_1 = 0.5$, $K'(A:B\parallel E)$ decreases as $\eta_2$ decreases (so when Eve gets more of the signal) but nonetheless, it remains positive, meaning that key exchange is always possible, albeit with reduced rate in the high loss regimes. 
Furthermore, unlike conventional continuous variable schemes, a central broadcast scheme has the advantage that it does not exhibit a sharp drop off, so a secret key can always be produced. 

\begin{figure}
\centering
\includegraphics[scale=0.4]{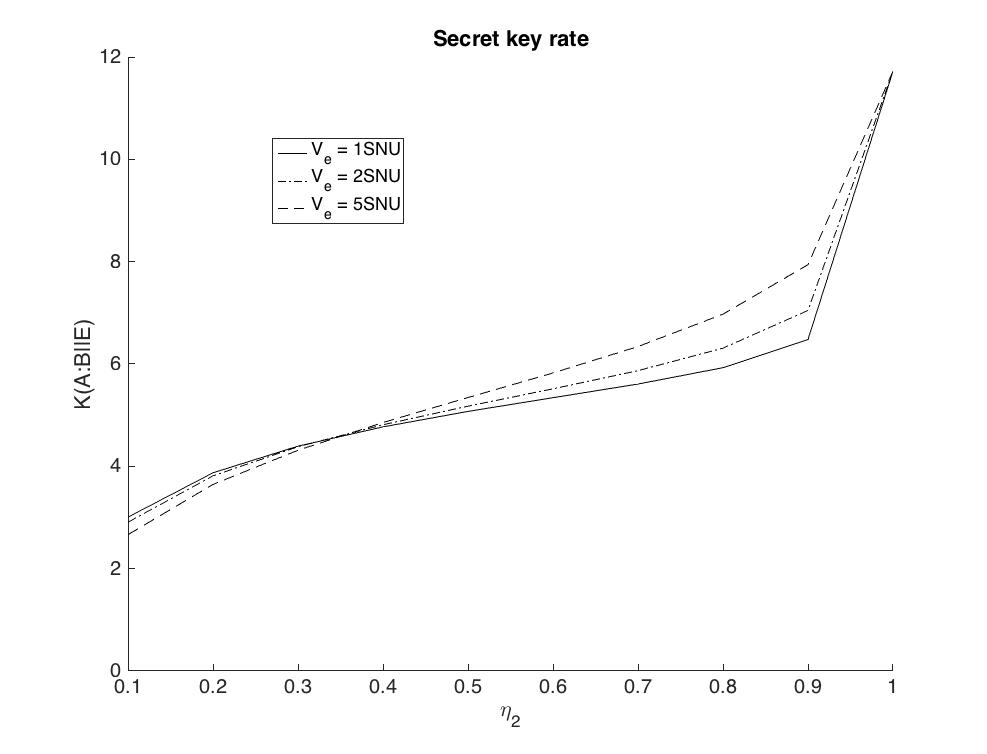}
\caption{(colour online) Secret key rate $K(A:B||E)$, plotted against $\eta_2$, for $\eta_4 = $ and $\eps = 10^{-2}$. The full line shows the secret key rate when the complementary input at $\eta_2$ is a vacuum state. The dashed-dotted line shows the secret key rate when Eve inputs a coherent state; the dashed line, when $V_e = 5 SNU$. Visibly, any input on the part of Eve improves the key rate. As $\eta_2$ approaches full transmission, the secret key rate increases rapidly. The x-axis is cropped for readability.}
\label{keyrates}
\end{figure}

As seen on Figure~\ref{keyrates}, the secret key rate is always positive, and so we conclude that there is always secrecy. 
It does, however, stagnate until $\eta_2$ approaches unity, i.e. until Eve lets most of the signal through. 
Key exchange is slow until Eve lets signal out; as a result, Eve is easily detectable. 
Let us recall also, that $K(A:B||E)$ is a lower bound; we can therefore expect better key rates. 
The secret key rate improves if Eve inputs a coherent state $V_e = 2SNU$, albeit marginally. 
This improvement is more pronounced if Eve inputs a thermal state $V_e = 5SNU$. 

\begin{figure}
\centering
\includegraphics[scale=0.4]{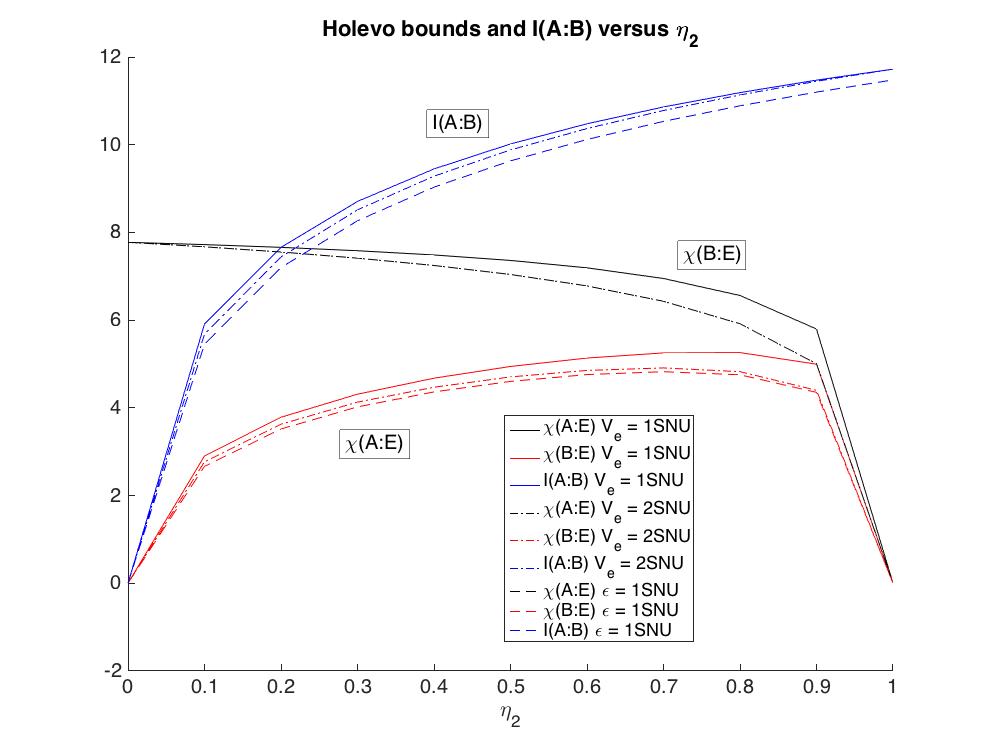}
\caption{(colour online) Holevo bounds and mutual information, plotted against $\eta_2$, for $\eta_4 = - 7dB $ and $\eps = 10^{-2}$. The full line indicate that the complementary input at $\eta_2$ is a vacuum state; the dashed-dotted line, that $V_e = 2SNU$.  Eve gains no more information by inputting a non-vacuum state; in fact, she fares worse. By contrast, the mutual information is only marginally reduced. The dashed line illustrates a channel excess noise $\eps = 1SNU$; as we expect, $I(A:B)$ and $\chi(B:E)$ suffer from a higher excess noise, but not $\chi(A:E)$ as neither Alice nor Eve are concerned with the thermal noise channel.}
\label{holmutinf}
\end{figure}

A crossover is obvious on the figure; at $\eta_2 = 30\%$ (so actually fairly low), Alice and Bob lose out if Eve inputs anything.
As $\eta_2$ increases, Bob gets more and more of the signal; as a result, Alice and Bob's mutual information increases, and as can be seen on Figure~\ref{holmutinf}, $I(A:B)$ increases faster than $\chi(B:E)$.

\begin{figure}
\centering
\includegraphics[scale=0.4]{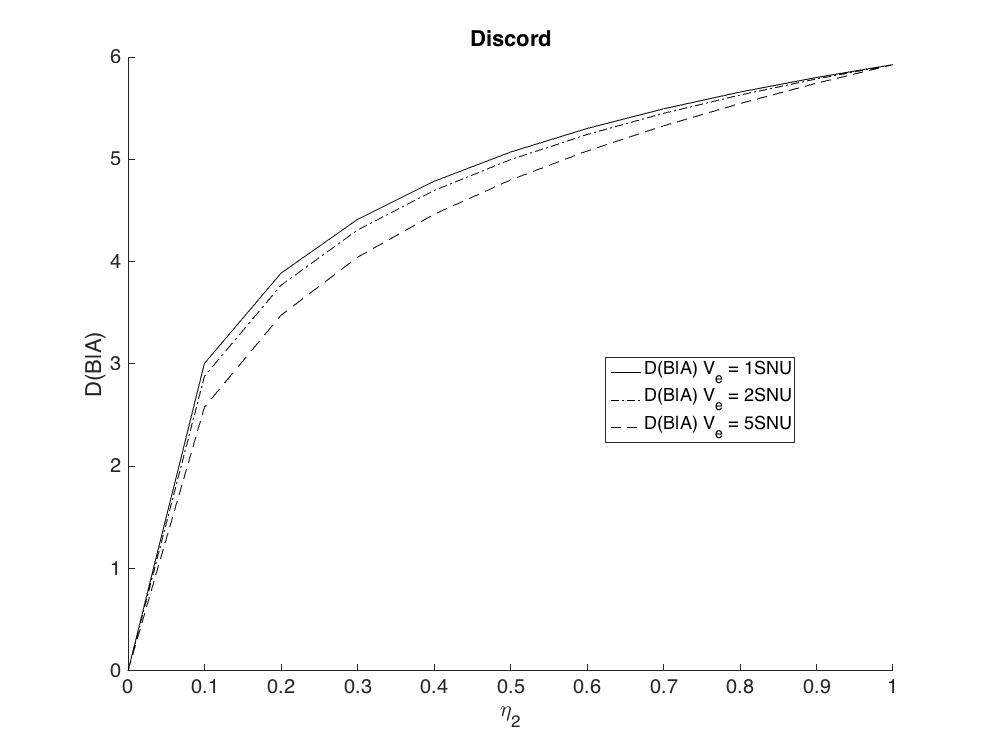}
\caption{(colour online) Discord $D(B|A)$, plotted against $\eta_2$, for $\eta_4 = - 7dB $ and $\eps = 10^{-2}$. The full line shows the discord when the complementary input at $\eta_2$ is a vacuum state. The dashed-dotted line shows the discord when Eve inputs a coherent state; the dashed line, when $V_e = 5 SNU$. As $\eta_2$ increases, so does the portion of the signal which goes to Bob and therefore, so does the discord. A non-vacuum complementary input at $\eta_2$ reduces the discord as it interferes with the correlated signal which Bob shares with Alice.}
\label{discords}
\end{figure}

Figure~\ref{discords} demonstrates the effect of the presence of Eve on the discord. 
Of highest importance to us is that the discord is always positive. 
This means that the correlations between Alice and Bob are always quantum, and as a result, so is our security. 
As the transmittance $\eta_2$ increases, Eve gets less and less signal, so naturally, the discord increases. 
When Eve's input is higher than 1SNU (so when she inputs a coherent state), the discord gets worse, but not significantly.

Figure~\ref{holmutinf} shows how Eve's input influences the amount of information she could gain, by the input of a state such as defined previously. 
Contrary to what one might expect, coming from a point-to-point prepare-and-send mindset, Eve does not gain much by inputting a coherent state. 
The Holevo bounds $\chi(A:E)$ and $\chi(B:E)$ are worse for $V_e = 2SNU$ than they are for $V_e = 1SNU$ (vacuum). 
This scheme relies on bunched radiation; Eve's effect is to destroys these pairs. 
As a result, she becomes (separately) correlated with Alice and with Bob, which is what the Holevo bounds show.
Her inputting a coherent state will not provide her with any more information, since it will not increase her correlation with either Alice nor Bob. 

On Figure~\ref{holmutinf}, we also explore briefly the influence of channel excess noise $\eps$ (dashed lines). 
We note that $\chi(A:E)$ is not affected by $\eps$; since the thermal noise channel is that which separates Eve and Bob, Alice is not affected.  
This is not the case for $I(A:B)$ and $\chi(B:E)$ which are both reduced by the channel excess noise. 
However, that reduction is insufficient to cause the key rate great damage.
It is reduced but not significantly enough to impede key distribution.

\section{Concluding remarks}

\sbp The outcome of this analysis is that our scheme can always be considered quantum secure, since both the secret key rate and the discord are positive. 
Furthermore, we have shown that Eve's input is merely a disturbance, not only to the legal parties but most importantly to herself. 
This is simply a result of the physics; the thermal radiation is correlated and since Eve can only place herself after it has been split at $\eta_1$ (and therefore she has no access to Alice's information), she can gain little information. 
Eve's attack might be construed as a jamming attack; she is detected prior to reconciliation, can construct no key, but she can prevent Alice and Bob from building key.

\sbp Although we have included thermal channel noise, we have so far neglected the issue of preparation noise and of detector noise. 
We note that any preparation noise from the source is thermal, and so a part of the thermal state input at $\eta_1$, therefore affects both Alice and Bob (and Eve). 

\begin{figure}
\centering
\includegraphics[scale=0.4]{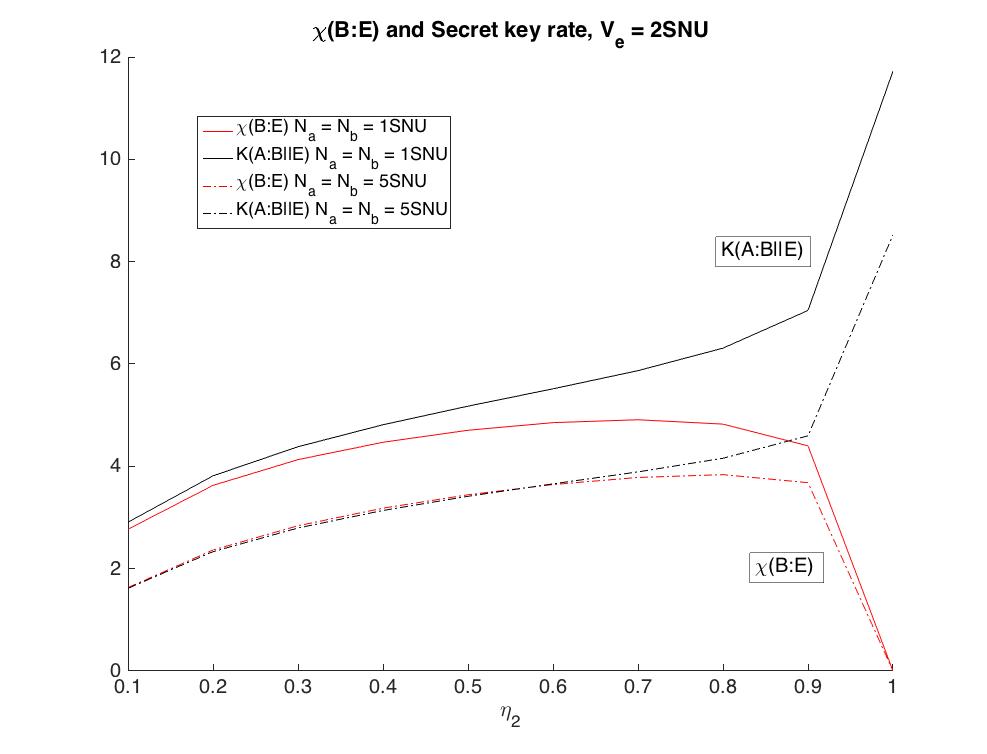}
\caption{(colour online) Effects of the detector noise on $\chi(B:E)$ and $K(A:B||E)$, $\eta_4 = - 7dB $ and $\eps = 10^{-2}$. The full lines indicates that the detector noise at Alice and Bob's detectors, respectively $N_a$ and $N_b$, are $1 SNU$. The dashed lines show such noise at $5 SNU$. The secret key rate suffers from detector noise, but so does $\chi(B:E)$. The x-axis is cropped for readability.}
\label{detectornoise}
\end{figure}

Detector noise, on the other hand, affects Alice and Bob individually and independently.
Adding detector noise to Alice's and Bob's signal yields $V_a = V_a^{pure} + N_a$ and $V_b = V_b^{pure} + N_b$.
We assume that Eve has no detector noise.
Figures~\ref{keyrates}, \ref{discords} and \ref{holmutinf} were plotted for detector noise at the mandatory unit of shot noise, $N_a = N_b = 1SNU$.
Detector noise will prevent Alice and Bob from detecting their correlations, but will not increase the correlations between Alice (Bob) and Eve. 
Therefore, it reduces the secret key rate, but also the Holevo bounds $\chi(A:E)$ and $\chi(B:E)$, as illustrated on Figure~\ref{detectornoise}.
Although the key rate remains positive, detector noise is the greatest threat to it.

\sbp There remains to be had perhaps, a discussion about distance. 
This issue is a subtle one.
Bunched pairs (and quantum correlations) survive astronomical distances; in fact, they only vanish upon measurement.
However, the correlations themselves will disappear if the photons are detected outside the coherence time. 
This means that the distance between $\eta_1$ and either Alice or Bob itself is not important; however, the path from $\eta_1$ to Alice and from $\eta_1$ to Bob should be of more or less equal length, i.e within the coherence time of the source. 
This of course, places restrictions on Eve's attack, such as we have already described.

\sbp A central broadcast scheme is attractive because the source need not be controlled to the extent of a point-to-point scheme, and two parties can negotiate a key based on their correlated local noise. 
This leads to a number of potential applications for key exchange in a microwave system, such as long distance satellite communications and for example, between mobile phones in a mobile network, mass synchronisation of secure keys within an office space connected to a WiFi node, or even key synchronisation between a mobile phone and an implanted medical device.

The authors are grateful to network collaborators J. Rarity, S. Pirandola, C. Ottaviani, T. Spiller, N. Luktenhaus and W. Munro for very fruitful discussions. This work was supported by funding through the EPSRC Quantum Communications Hub EP/M013472/1 and additional funding for F.W. from Airbus Defense \& Space.

\bibliographystyle{unsrt}

\bibliography{cvqkd_bibli}


\appendix

\section{Full results \label{prot1submatrices}}
On the output of $\eta_2$, the submatrices are as follows
{\alld
\begin{align}
\gm_a =& \left( \begin{array}{cc} \mu_1^2 V_s^x + \eta_1 & 0 \\ 0 &\mu_1^2 V_s^p + \eta_1 \end{array} \right)\,, \non
\\ \gm_b =& \left( \begin{array}{cc} \eta_2 (\eta_1 V_s^x + \mu_1^2) + \mu_2^2V_e^x  & 0 \\ 0 & \eta_2 (\eta_1 V_s^p + \mu_1^2) + \mu_2^2 V_e^p \end{array} \right)\,, \non
\\ \gm_e =& \left( \begin{array}{cc}  \mu_2^2 (\eta_1 V_s^x + \mu_1^2) + \eta_2 V_e^x & 0 \\ 0 & \mu_2^2 (\eta_1 V_s^p + \mu_1^2) + \eta_2 V_e^p  \end{array} \right)\,, \non
\\ \gm_{ea} =& \left( \begin{array}{cc}  \mu_1 \sqrt{\eta_1} \mu_2 (V_s^x-1) & 0 \\ 0 & \mu_1 \sqrt{\eta_1} \mu_2  (V_s^p-1) \end{array} \right)\,, \non \\ \gm_{eb} =& \left(\begin{array}{cc} - \mu_2 \sqrt{\eta_2} (\eta_1 V_s^x + \mu_1^2 -V_e^x) & 0 \\ 0 & - \mu_2 \sqrt{\eta_2} (\eta_1 V_s^p + \mu_1^2-V_e^p  ) \end{array} \right) \,, \non 
\\   \gm_{ab} =& \left(\begin{array}{cc} - \mu_1 \sqrt{\eta_1} \sqrt{\eta_2} (V_s^x-1) & 0 \\ 0 & - \mu_1 \sqrt{\eta_1} \sqrt{\eta_2}  (V_s^p-1) \end{array} \right) \non
\end{align}}

On the output of $\eta_4$, the submatrices are as follows
{\alld
\begin{align}\Gm_a =& \left( \begin{array}{cc} \expvc{X_a^2} & 0 \\ 0 & \expvc{P_a^2} \end{array} \right)\,, \quad \Gm_e = \left( \begin{array}{cc} \expvc{X_e^2} & 0 \\ 0 & \expvc{P_e^2} \end{array} \right)\,, \non
\\ \Gm_{ea} =& \left( \begin{array}{cc} \expvc{X_a X_e} & 0 \\ 0 & \expvc{P_a P_e} \end{array} \right)\,, \quad \Gm_{eb} = \sqrt{\eta_4} \gm_{eb} \,, \non
\\ \Gm_{ev} =& -\mu_4 \gm_{eb} \,, \quad \Gm_{ab} = \sqrt{\eta_4} \gm_{ab} \,, \quad \Gm_{an} = -\mu_4 \gm_{ab} \non 
\\ \Gm_b =& \left( \begin{array}{cc} \eta_4 \expvc{X_b^2} + \mu_4^2 N & 0 \\ 0 &  \eta_4 \expvc{P_b^2} + \mu_4^2 N \end{array} \right) \,, \non
\\\Gm_n =& \left(\begin{array}{cc} \mu_4^2 \expvc{X_b^2} + \eta_4 N & 0 \\ 0 &  \mu_4^2 \expvc{P_b^2} + \eta_4 N \end{array} \right) \non \\ \Gm_{bn} =& \left(\begin{array}{cc} \mu_4 \sqrt{\eta_4}\left(N- \expvc{X_b^2} \right)  & 0 \\ 0 & \mu_4 \sqrt{\eta_4}\left(N - \expvc{P_b^2} \right)  \end{array} \right) \non
\end{align}}

\end{document}